\documentclass[aps,prd,onecolumn,showpacs,nofootinbib,amsmath,amssymb,floatfix,superscriptaddress,showkeys]{revtex4}
\usepackage{graphicx}% Include figure files
\usepackage{dcolumn}% Align table columns on decimal point
\usepackage{bm}% bold math
\usepackage{captcont}
\usepackage{xcolor}
\usepackage{supertabular}

\usepackage{epstopdf}
\usepackage{mathtools}
\usepackage{natbib}
\usepackage{ulem}
\usepackage[colorlinks,linkcolor=red,anchorcolor=blue,citecolor=blue]{hyperref}

\newcommand{\jcap}{J. Cosmol. Astropart. Phys.}
\newcommand{\apjl}{Astrophys. J. Lett.}
\newcommand{\physrep}{Phys. Rep.}
\newcommand{\mnras}{Mon. Not. R. Astron. Soc.}
\newcommand{\araa}{Annu. Rev. Astron. Astrophys.}

\newcommand{\aj}{Astron. J.}
\newcommand{\plb}{Phys. Lett. B}

\begin{document}

\title{3FGL J1924.8-1034 : A Spatially-Extended Stable Unidentified GeV Source?}

\author{Zi-Qing Xia}
\affiliation{Key Laboratory of Dark Matter and Space Astronomy, Purple Mountain Observatory, Chinese Academy of Sciences, Nanjing 210008, China}
\affiliation{School of Astronomy and Space Science, University of Science and Technology of China, Hefei, Anhui 230026, China}
\author{Kai-Kai Duan}
\affiliation{Key Laboratory of Dark Matter and Space Astronomy, Purple Mountain Observatory, Chinese Academy of Sciences, Nanjing 210008, China}
\affiliation{University of Chinese Academy of Sciences, Beijing, 100012, China}
\author{Shang Li}
\email{lishang@pmo.ac.cn}
\affiliation{Key Laboratory of Dark Matter and Space Astronomy, Purple Mountain Observatory, Chinese Academy of Sciences, Nanjing 210008, China}
\affiliation{University of Chinese Academy of Sciences, Beijing, 100012, China}
\author{Yun-Feng Liang}
\email{liangyf@pmo.ac.cn}
\affiliation{Key Laboratory of Dark Matter and Space Astronomy, Purple Mountain Observatory, Chinese Academy of Sciences, Nanjing 210008, China}
\affiliation{University of Chinese Academy of Sciences, Beijing, 100012, China}
\author{Zhao-Qiang Shen}
\email{zqshen@pmo.ac.cn}
\affiliation{Key Laboratory of Dark Matter and Space Astronomy, Purple Mountain Observatory, Chinese Academy of Sciences, Nanjing 210008, China}
\affiliation{University of Chinese Academy of Sciences, Beijing, 100012, China}
\author{Chuan Yue}
\affiliation{Key Laboratory of Dark Matter and Space Astronomy, Purple Mountain Observatory, Chinese Academy of Sciences, Nanjing 210008, China}
\affiliation{University of Chinese Academy of Sciences, Beijing, 100012, China}
\author{Yuan-Peng Wang}
\affiliation{Key Laboratory of Dark Matter and Space Astronomy, Purple Mountain Observatory, Chinese Academy of Sciences, Nanjing 210008, China}
\affiliation{University of Chinese Academy of Sciences, Beijing, 100012, China}
\author{Qiang Yuan}
\affiliation{Key Laboratory of Dark Matter and Space Astronomy, Purple Mountain Observatory, Chinese Academy of Sciences, Nanjing 210008, China}
\affiliation{School of Astronomy and Space Science, University of Science and Technology of China, Hefei, Anhui 230026, China}
\author{Yi-Zhong Fan}
\email{yzfan@pmo.ac.cn}
\affiliation{Key Laboratory of Dark Matter and Space Astronomy, Purple Mountain Observatory, Chinese Academy of Sciences, Nanjing 210008, China}
\affiliation{School of Astronomy and Space Science, University of Science and Technology of China, Hefei, Anhui 230026, China}
\author{Jian Wu}
\affiliation{Key Laboratory of Dark Matter and Space Astronomy, Purple Mountain Observatory, Chinese Academy of Sciences, Nanjing 210008, China}
\affiliation{School of Astronomy and Space Science, University of Science and Technology of China, Hefei, Anhui 230026, China}
\author{Jin Chang}
\affiliation{Key Laboratory of Dark Matter and Space Astronomy, Purple Mountain Observatory, Chinese Academy of Sciences, Nanjing 210008, China}
\affiliation{School of Astronomy and Space Science, University of Science and Technology of China, Hefei, Anhui 230026, China}

\date{\today}

\begin{abstract}
Milky Way-like galaxies are predicted to host a very large number of dark matter subhalos. Some massive and nearby subhalos could generate detectable gamma-rays, appearing as unidentified, spatially-extended and stable gamma-ray sources. We search for such sources in the third Fermi Large Area Telescope source List (3FGL) and report the identification of a new candidate, 3FGL J1924.8-1034. With the Fermi-LAT Pass 8 data, we find that 3FGL J1924.8-1034 is spatially-extended at a high confidence level of $5.4\sigma$, with a best-fit extension radius of $\sim0.15^{\circ}$.  No significant variability has been found and its gamma-ray spectrum is well fitted by the dark matter annihilation into $b\bar{b}$ with a mass of $\sim 43$ GeV. All these facts make 3FGL J1924.8-1034 a possible dark matter subhalo candidate. However, due to the limited angular resolution, the possibility of that the spatial extension of 3FGL J1924.8-1034 is caused by the contamination from the other un-resolved point source can not be ruled out.
\end{abstract}


\maketitle

\section{Introduction}
The total mass$-$energy of the universe can not be explained by the ordinary matter in the standard model of cosmology. Dark matter, an invisible form of matter, is believed to make up a quarter of the energy density of the current Universe. However, nobody knows what it is. Various hypothetical particles have been proposed. As the most widely accepted hypothesis, it is composed of weakly-interacting massive particles (WIMPs), which only interact through gravitational and weak forces and may be able to annihilate with each other (or alternatively decay) into stable high$-$energy particle pairs, including gamma rays, charged particles and neutrinos \cite[]{wimp1,wimp2,wimp3,wimp4,wimp5}. The identification of these annihilation or decay products is the main goal of dark matter indirect detection experiments.

Among various indirect search methods, observations of the gamma-ray sky have attracted wide attention during these years thanks to the successful and productive performance of the Fermi Large Area Telescope (LAT) \cite{charles16review,fermi2009}. Intense dedicated researches have been carried out to catch the gamma-ray radiation from dark matter annihilation or decay utilizing the Fermi-LAT data towards many different observational targets, such as the dwarf spherical galaxies \cite{fermi11dsph,fermi14dsph,fermi15dsph,hooper15ret2,gs15ret2,des_fermi15dsph,gs15dsph,li16dsph,fermi2016dsph}, galaxy clusters \cite{huang12cluster,Yuancluster,ando12fornax,fermi15virgo,Liang2016}, the Galactic center \cite[]{Ajello2016,hooper2011,gordon13gc,hooper13gc,zb15gc,calore15gc,huanggc}, and the extragalactic gamma-ray background \cite[]{isoFermi,isoMF,isoLW}. Among these various targets, the most promising are the Galactic center for its high density of dark matter and the dwarf spherical galaxies (dSphs) for the low astrophysical gamma$-$ray backgrounds. However so far no reliable dark matter signal has been identified \cite{charles16review}. The joint analysis of fifteen dwarf spheroidal galaxies sets the most stringent constraint on cross section of DM annihilating through $b\bar{b}$ and $\tau^+\tau^-$ channels \cite{fermi15dsph}.

In this work, we focus on dark matter subhalos. N-body simulations reveal that dark matter structures form hierarchically, i.e.,  the dark matter gathers together to form small halos, and then small halos merge repeatedly to create ever larger systems \cite{nfw}. As a consequence of this process, a Milky Way-like galaxy is predicted to host tens of thousands of Galactic DM subhalos \cite{DMre145,DMre182}. The most massive ones of these subhalos are expected to host the known dSphs, while some small subhalos may completely lack any astrophysical counterparts for no significant quantity of baryonic matter. However, some of these small subhalos, if still massive enough and close enough to the Earth, may generate detectable gamma-rays, observed as a group of  unidentified gamma$-$ray sources. For example, assuming a dark matter particle mass of $\sim 100$ GeV and an annihilation cross section of ${\rm \langle \sigma v \rangle \simeq 2\times10^{-26}~cm^3~s^{-1}}$, Fermi-LAT might have recorded $\sim 10$ dark matter subhalos \cite{dan2015sh,sh1,sh2,sh3,sh4,sh5}. To further distinguish between the unidentified astrophysical sources and the dark matter subhalos, the spatial extension likely plays an important role since a spatially-extended stable source without any association in other wavelengths is hard to be explained in astrophysical scenarios \cite{dan2016sh1}. Consequently, we consider the unidentified stable gamma-ray sources with extendible structure at a confidence level of $>5 \sigma$ (${\rm \Delta TS_{\rm ext}> 25}$) as DM subhalo candidates.

The rest of this paper is structured as follows: In the second section, we present our selection of DM subhalo candidates and the comparison with the previous literature. In the third section, we describe details of the Fermi data analysis focusing on one single DM subhalo candidate. Finally, in the fourth section, we summarize our results and discuss some uncertainties in our work. 

\section{ THE SELECTION OF  DM SUBHALO CANDIDATES}

There are 3033 sources included in the third Fermi Large Area Telescope source List (3FGL) \cite{3FGL}. 992 of them have not been identified in other wavelengths.  In our investigation, we searched for dark matter subhalo candidates among the unidentified stable (with Variability\_Index less than 80) sources at galactic latitudes ${\rm |b|>10^{\circ}}$ without any cut in gamma-ray flux intensity in the 3FGL. The spatial characteristics of all these sources are studied in our work. We find that 3FGL J1924.8-1034 is the only source which passes the spatial extension test at a confidence level $>5\sigma$ and  3FGL J2212.5+0703 is also spatially-extended at a relatively lower confidence level of $4.7\sigma$.

Bertoni et al. \cite{dan2015sh} have analyzed the unidentified, stable and bright (i.e., with the gamma-ray flux $F_\gamma >7\times10^{-10}~{\rm cm^{-2}~s^{-1}}$ when integrated above 1 GeV) GeV sources at high galactic latitudes ($|b|>5^{\circ}$) and found that 3FGL J2212.5+0703 stands out as a subhalo candidate but 3FGL J1924.8-1034 does not.
To understand the difference we'd like to compare the difference between the data processing procedures. Bertoni et al. \cite{dan2015sh} didn't fit the best position of 3FGL J1924.8-1034 by themselves, instead they adopted the position reported in 3FGL that was based on 4 years Fermi-LAT data. In this work, the best position fitted with the latest data (deviating from the 3FGL position by $\sim 0.1$ deg) is adopted in the following analysis. The other difference is that we adopt Pass 8 data while just Pass 7 data was available in the previous study.

Since 3FGL J2212.5+0703 has already been discussed in great detail in literature \cite{dan2015sh,dan2016sh1,wang2016}, in the rest work we focus on the unidentified gamma-ray source 3FGL J1924.8-1034 and analyze the spatial distribution, light curve and spectrum of the gamma-ray emission from this source. The possibility of that the spatial extension of 3FGL J1924.8-1034 is due to the contamination of a nearby  un-resolved point source has also been examined.

\section{FERMI DATA ANALYSIS}
\label{samples}
To study the variability, spectrum and morphology of 3FGL J1924.8-1034, we use 94 months of Pass 8 LAT data from 2008 Aug 27 (MET = 239500801s) to 2016 Jun 01 (MET = 486432004s) in the SOURCE event class with the standard conversion-type (FRONT+BACK) selection. Photons with energy range between 300 MeV and 300 GeV are taken into consideration. 3FGL J1924.8-1034  at the galactic latitudes (${\rm b = -12.08^{\circ}}$) is close to the galactic disk, and the photons below 300 MeV is  contaminated by the diffuse galactic  emission. The zenith angle cut ${\rm \theta} < 90^{\rm \circ}$ is applied to reduce the contribution from Earth Limb and we adopt the recommended quality-filter cuts (DATA\_QUAL{\textgreater}0 and LAT\_CONFIG==1) to extract the good time intervals. Then we create $14^{\rm \circ} \times 14^{\rm \circ}$ regions of interest (ROI) centered on (RA, DEC, J2000) = ($291.21^{\rm \circ},~-10.58 ^{\rm \circ}$) \cite{3FGL} and perform a standard binned likelihood analysis with $0.05^{\rm \circ}$ spatial bins and 30 logarithmic energy bins. Fermi Science Tools v10r0p5 and instrument response functions (IRFs) P8R2\_SOURCE\_V6 are used for this analysis, which are available from the Fermi Science Support Center.\footnote{http://fermi.gsfc.nasa.gov/ssc/}

Utilizing the user-contributed script make3FGLxml.py \footnote{http://fermi.gsfc.nasa.gov/ssc/data/analysis/user/}, we make our model as a combination of the latest model for diffuse Galactic gamma-ray emission (gll\_iem\_v06.fits), the latest isotropic emission for the SOURCE photon data selection (iso\_P8R2\_SOURCE\_V6\_v06.txt)  and all 3FGL sources  found within a $20^{\rm \circ}$  of our region of interest (ROI). We free the normalizations and spectral indexes of all sources within $7^{\rm \circ}$  from the target source and the normalizations of the two diffuse emission backgrounds in the maximum likelihood fit.We take the {\it PowerLaw} as the spectral shape of 3FGL J1924.8-1034, which is the default setting of 3FGL J1924.8-1034 in the 3FGL \cite{3FGL}. To perform the fit and get an optimized model, we use the Fermi-LAT $ pyLikelihood$ code, utilizing the {\it MINUIT} algorithm \cite{minuit}.

\begin{figure}%[!h]
\includegraphics[width=0.48\columnwidth]{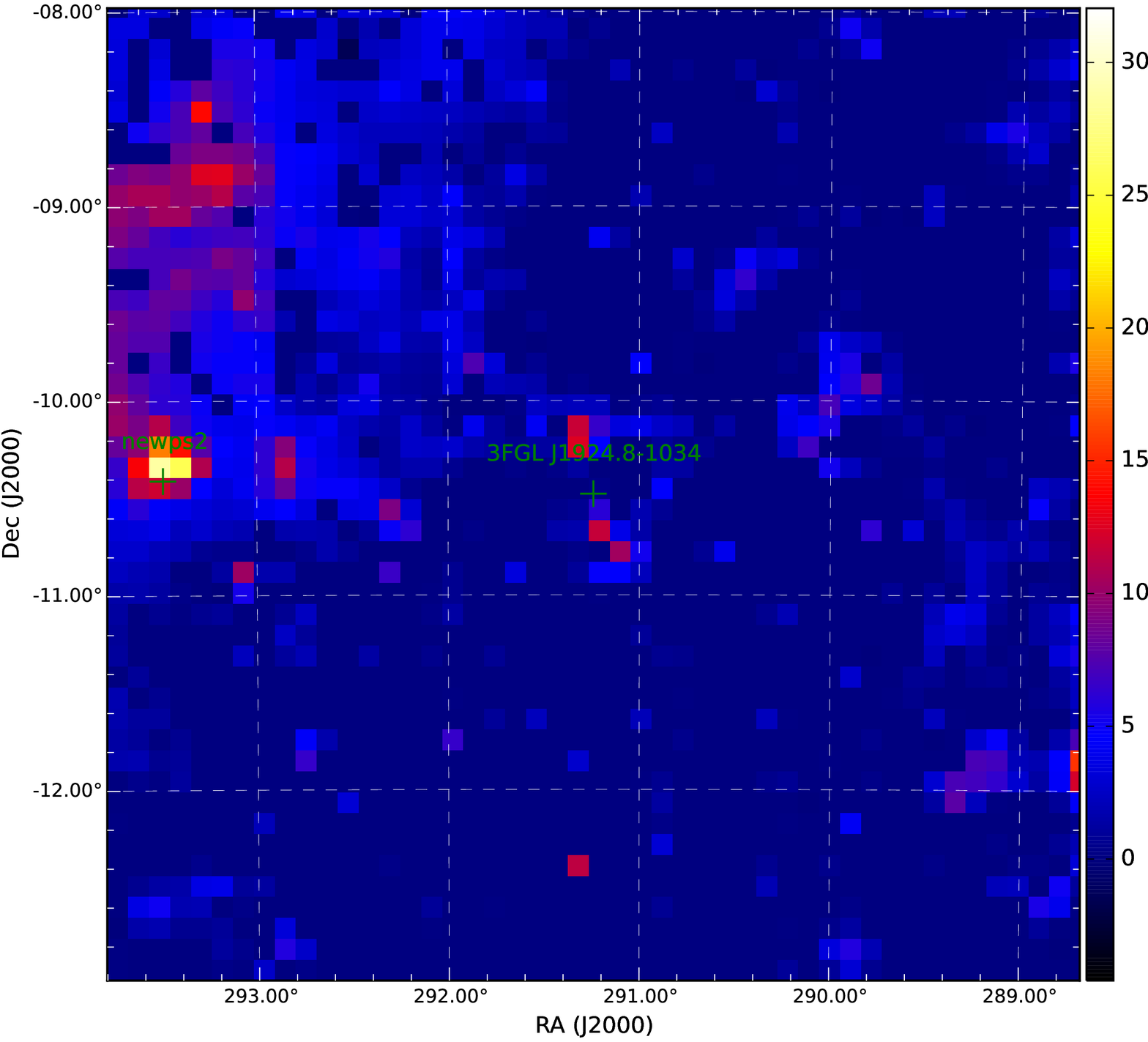}
\includegraphics[width=0.48\columnwidth]{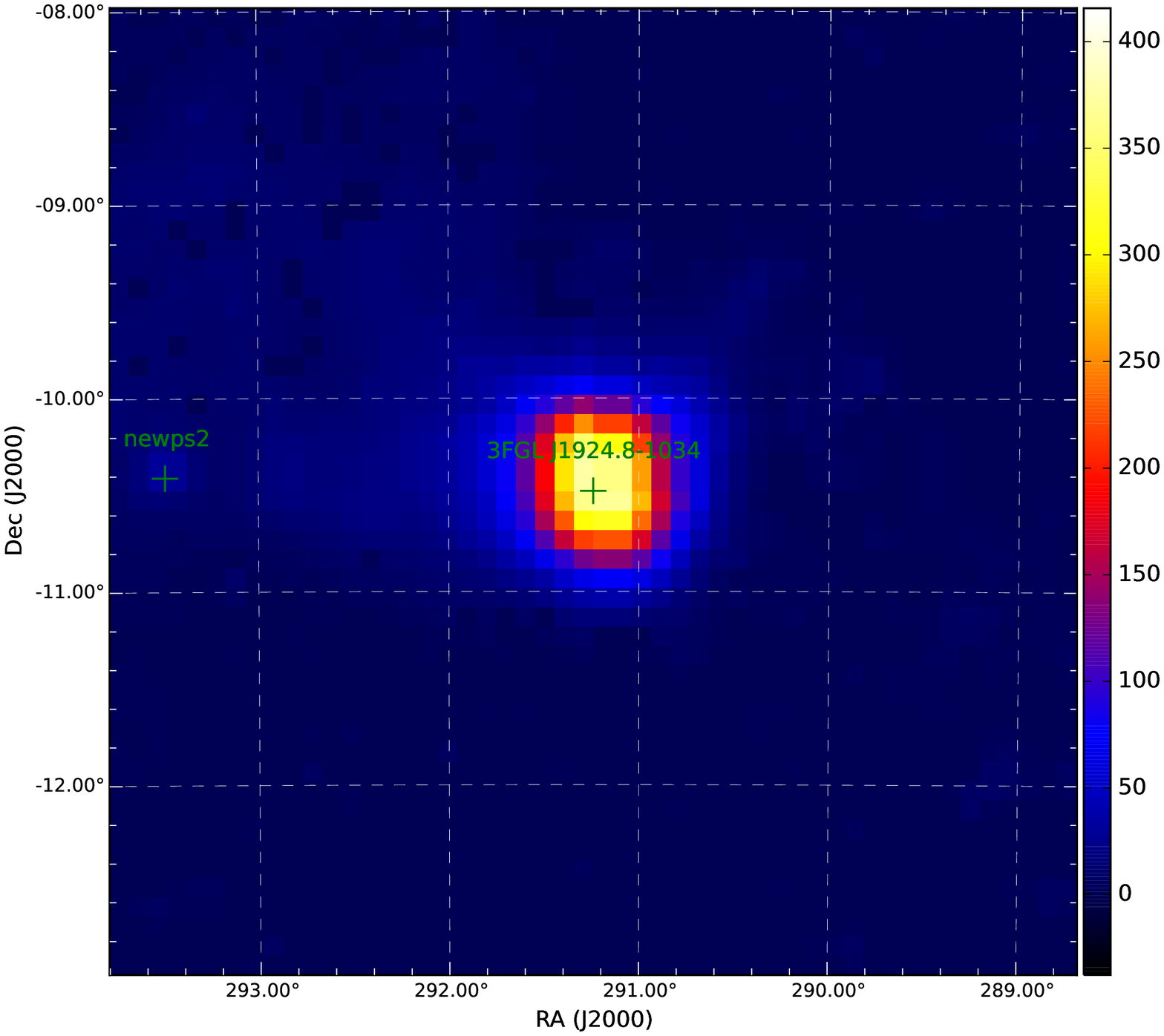}
\caption{$5^{\rm \circ} \times 5^{\rm \circ}$ TS map centered at the position of 3FGL J1924.8-1034. In the left (right) panel we show the TS map with (without) a point target source included in the modeling. The green crosses represent the positions of the sources within $5^{\rm \circ} \times 5^{\rm \circ}$ listed in Tab.\ref{tb1}}
\label{tsmap}
\end{figure}

We derive  a $10^{\circ} \times 10^{\circ}$ Test Statistic (TS) map by placing a test point source at the location of each pixel of the map and maximizing the likelihood which is implemented in the $gttsmap$ tool \footnote{http://fermi.gsfc.nasa.gov/ssc/data/analysis/scitools/help/gttsmap.txt}. The TS is defined as ${\rm TS = -2ln(}\mathcal{L}_{\rm max,0}/\mathcal{L}_{\rm max,1})$ following \cite{TS1996}, where $\mathcal{L}_{\rm max,0}$ is the maximum likelihood value for a model without an additional source (the null hypothesis) and $\mathcal{L}_{\rm max,1}$ is for a model with that additional source(the alternative hypothesis). We find two new point sources with a TS value larger than 25 and add them to our model with power-law spectra located at the corresponding positions on the TS map.  With the help of {\it gtfindsrc}, We find the optimized positions of  the target source 3FGL J1924.8-1034 and two new added sources shown in Tab.\ref{tb1}. {The $5^{\rm \circ} \times 5^{\rm \circ}$ TS map with (without) a point target source is displayed in Fig.\ref{tsmap}.}

\begin{table}[ht]
\small
\begin{tabular}{lrrcccccc}
\hline
\hline
Source Name & R.A.[$^{\rm \circ}$] & Decl.[$^{\rm \circ}$]  \\
\hline
3FGL J1924.8-1034  & 291.24  &  -10.48 \\
newps1  & 288.20  &  -12.83 \\
newps2  & 293.49  &  -10.41 \\
\hline
\end{tabular}
\caption{The optimized positions of  the target source and two new additional sources.}
\label{tb1}
\end{table}

\subsection{The Spatial Extension}
To determine whether 3FGL J1924.8-1034 exhibits any evidence of spatial extension, we replace the point-source template with a series of extended templates varying the width as a free parameter \cite{dan2015sh,wang2016}. We measure the width by the parameter ${\theta_{68}}$, defined as the angular radius which contains 68\% of the total photons from the target source.

\begin{figure}%[!h]
\includegraphics[width=0.6\columnwidth]{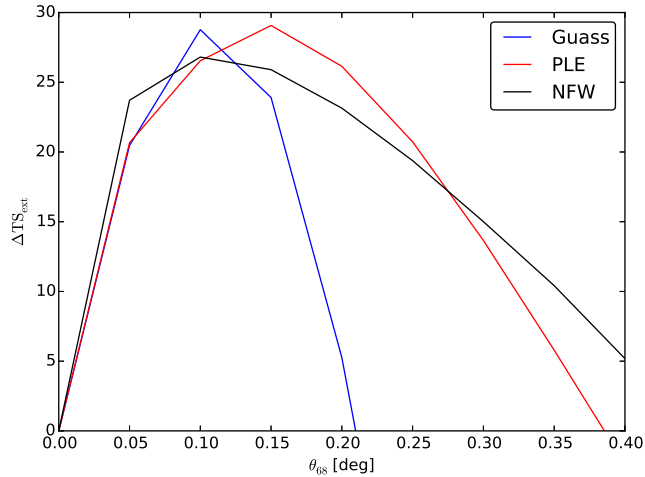}
\caption{The ${\rm \Delta TS_{\rm ext}}$ of 3FGL J1924.8-1034 when the point-like template is replaced with a spatial template in the cases of three extension profiles (GAUSS, NFW, PLE). The (GAUSS, NFW, PLE) template fit peaks at $\theta_{68}=(0.1^{\rm \circ},~0.1^{\rm \circ},~0.15^{\rm \circ})$ with a ${\rm \Delta TS_{\rm ext}=(28.7,~26.8,~29.1)}$, respectively.}
\label{ext}
\end{figure}

 A spatially-extended stable source without any association in other wavelengths is hard to be explained by astrophysical origin. Instead it is expected for the massive or nearby dark matter subhalo \cite{dan2016sh1}. The density in the dark matter host halo is often considered as Navarro-Frenk-White (NFW) profile \cite{nfw1,nfw2}. However, the shapes of nearby subhalos,  significantly altered by tidal effects, are generally not well described by NFW density profiles. Instead they prefer NFW profiles with an exponential cutoff \cite{Ghigna98,Kazantzidis04,Diemand07a,Diemand07b,Moline16} or power-law profiles with an exponential cutoff (PLE) \cite{dan2016sh2}.

For completeness, we choose the three kinds of distribution of the extension: (1)The 2-D Gaussian distribution (GAUSS) \cite{dan2015sh}, (2) A distribution corresponding to a NFW dark matter density profile \cite{dan2016sh1},  (3) A distribution corresponding to a PLE dark matter density profile \cite{dan2016sh2}.

The NFW density profile is given by \cite{nfw1,nfw2}
\begin{equation}
{ \rho(r)}= \frac{ \rho_0}{( r/R_s)[1+( r/R_s)]^2}.
\end{equation}
 where ${R_s}$ is the subhalo's scale radius. We adopt the mass-concentration model presented in the \cite{Moline16} for subhalos.

The PLE density profile is given by
\begin{equation}
{ \rho(r)}= \frac{ \rho_0}{r^\gamma}\exp(-{r \over R_{\rm b}}),
\end{equation}
where the index ${\gamma}$ is taken as 0.74 and  $R_{\rm b}$ is the cutoff radius\cite{dan2016sh2} .

The extended template corresponding to dark matter density ${ \rho(r)}$ (where ${r({\theta, d, l})=\sqrt{d^2+l^2-2dl\cos\theta}}$ ) is governed by
\begin{equation}
{ f(\theta)}= \frac{\int_{l=0}^{ \infty}{\rho^2(r({\theta,d,l}))}{dl}}{ \int_0^{\theta_{\rm max}}2\pi\theta\int_{l=0}^\infty{\rho^2(r(\theta,d,l))}{dl}{d\theta}},
\end{equation}
where the first integrals are performed over the line-of-sight, ${\theta}$ is the angle to the center of the subhalo,  ${\theta_{\rm max}}$  is the angular radius encompassing the full extension of the subhalo, and $d$ is the distance to the center of the subhalo. The width (${\theta_{68}}$) for this extended template can be defined as ${\int_ 0^{\theta_{68}}2\pi\theta\times f(\theta){d\theta}= 68 \%}$.

In order to compare these spatial templates, we do the likelihood fit with each template. The TS for spatial extension is defined as the change to the log-likelihood, when the point-like template is replaced with that of an extended source:
\begin{equation}
{\rm \Delta TS_{\rm ext}}= -2(\ln{\mathcal{L}_{\rm point}}-\ln{\mathcal{L}_{\rm ext }}).
\end{equation}
where ${\mathcal{L}_{\rm point}}$ and ${\mathcal{L}_{\rm ext}}$ are the best fit likelihood values for the point-source model and the extended model.

The results plotted in Fig.\ref{ext}, imply that 3FGL J1924.8-1034 prefers a spatially-extended profile over that of a single point-like source at a level of ${> 5 \sigma}$ (${\rm \Delta TS_{\rm ext}> 25}$) for all three kinds of extended templates and the PLE template with a width of $0.15^{\rm \circ}$  best describes the data with ${\rm \Delta TS_{\rm ext}}$ = 29.1, corresponding to a statistical significance of ${~ 5.4\sigma}$. Our results are insensitive on the adopted extension template.

 Considering that the Fermi LAT point spread function(PSF) is large below 3GeV, we restrict the energy range from 3 GeV to 300 GeV to reselect photons and perform the same analysis. We still find that 3FGL J1924.8-1034 prefers a GAUSS-extended profile with a width of $0.1^{\rm \circ}$ than that of a single point-like source at a confidence level of ${ 4.2 \sigma}$ (${\rm \Delta TS_{\rm ext} =17.9}$). Moreover we analyze the PSF3-event-type Pass 8 data which has the best angular resolution for a given energy and get a ${\rm \Delta TS_{\rm ext}}=15.2$ for a $0.1^{\rm \circ}$-width GAUSS template.

 In the above analysis, we choose the spherically symmetric templates to study the spatial extension of 3FGL J1924.8-1034. According to the numerical simulations, whereas isolated dark matter halos tend to be spherically, subhalos are predominantly triaxial \cite{Kuhlen07,Carlos14}.  So we change the extension template to the spherically symmetric GUASS template stretched along one axis and free the width, stretch ratio and orientation to repeat the likelihood fit. The Stretched template, with a width of $0.1^{\rm \circ}$,  a stretch ratio of 1.3 and a stretched axis oriented approximately $14^{\rm \circ}$  counterclockwise from the vertical direction, provides the best fit with ${\rm \Delta TS_{\rm ext} =43.8}$ corresponding to a statistical significance of ${~ 6.0\sigma}$ for three additional degrees-of-freedom. The Stretched template is preferred over the best-fit spherically symmetric template at a level of ${\rm \Delta \ln{\mathcal{L}=14.7}}$ with two additional degrees-of-freedom.

However, the Fermi LAT diffuse model is far from being perfect. Some small-size diffuse structures may be not present in this model, which may lead to biased conclusions. Such a possibility can not be ruled out for the current source.

\subsection{The Variability}
\begin{figure}%[!h]
\includegraphics[width=0.6\columnwidth]{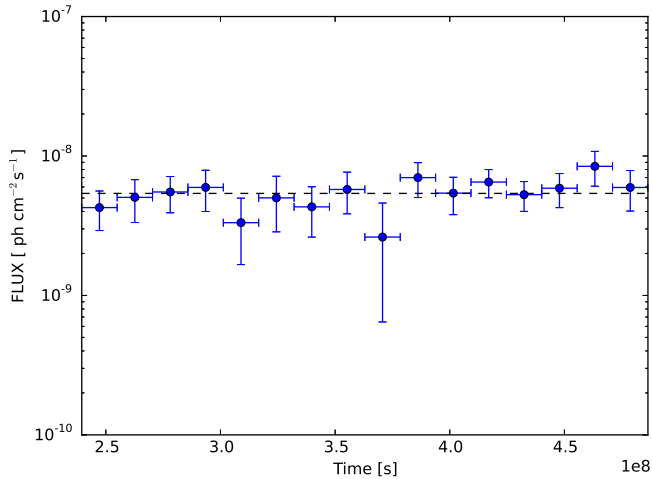}
\caption{The light curve of 3FGL J1924.8-1034. The blue point is the photon flux integrated from
300 MeV to 300 GeV in each time bin, and the dashed line is the mean value of all blue points.}
\label{lc}
\end{figure}

Considering that 3FGL only used 4 years Fermi LAT data, there is a need to check whether the signal from 3FGL J1924.8-1034 has been stable for almost 8 years \citep[see also][]{wang2016}.  To calculate the light curve of 3FGL J1924.8-1034, we divide these photons into 16 equal time bins. We fix all the parameters except the normalizations of the target source and the diffuse backgrounds in the optimized model. The likelihood fits are performed with this modified model independently in each time bin. The light curve of 3FGL J1924.8-1034 is shown in Fig.\ref{lc}. To test the variability of 3FGL J1924.8-1034,  a variability index is constructed as \cite{2FGL,wang2016},

\begin{equation}
{\rm TS_{\rm var}}= -2\sum_{\rm i}\frac{ \Delta F_{\rm i}^2}{ \Delta F_{\rm i}^2+f^2F_{\rm const}^2}\ln\frac{\mathcal{L}_{\rm i}({ F_{\rm const} })}{\mathcal{L}_{\rm i}({ F_{\rm i} })}.
\end{equation}
${\mathcal{L}_{\rm i}({ F_{\rm const} }) }$ is the value of the likelihood in the i$-$th bin under the null hypothesis where the source flux is constant across the full period and $F_{\rm const}$ is the constant flux for the this hypothesis, while ${\mathcal{L}_{\rm i}({ F_{\rm i} }) }$ is the value under the alternate hypothesis where the flux in the i$-$th bin is optimized. For each time bin, the photon flux over the full energy range (300 MeV to 300 GeV) is $F_{\rm i}$, and its statistical error is  $\Delta F_{\rm i}$. And we take $f=2\%$ as the systematic correction factor \cite{3FGL}.

If the null hypothesis is correct, ${\rm TS_{\rm var}}$ is distributed as ${\chi ^2}$ with 15 degrees of freedom \cite{wilk}. We find ${\rm TS}_{\rm var}=13$ and the alternative hypothesis has a low significance of 0.52$\sigma$. So the null hypothesis is reasonable and 3FGL J1924.8-1034 is an non-variable source.

\subsection{The Spectrum}
From the above analysis, we identify 3FGL J1924.8-1034 as a stable and spatially-extended gamma-ray source without any association in other wavelengths, which makes it a potential candidate of dark matter subhalo. We assume that the gamma-ray emission observed from 3FGL J1924.8-1034 was generated by DM annihilation. The gamma-ray intensity generated by DM annihilation in the DM subhalo can be described by
\begin{equation}
\Phi(E)=\frac{1}{4\pi}\frac{\langle\sigma{v}\rangle}{2m_{\chi}^2}\frac{dN}{dE}\cdot{J_{\rm factor}},
\end{equation}
where $\langle\sigma{v}\rangle$ is the annihilation cross section averaged by the DM velocity and $m_{\chi}$ is the rest mass of the DM particle.The  ${J_{\rm factor}}$ is the integration of the square of DM density $\rho(r)$ along the line of sight $s$, i.e.,
\begin{equation}
{J_{\rm factor} = \int_\Omega\int_{s=0}^\infty{\rho^2(r(s))}{\rm d}s{\rm d}\Omega}.
\end{equation}

\begin{figure}%[!h]
\includegraphics[width=0.6\columnwidth]{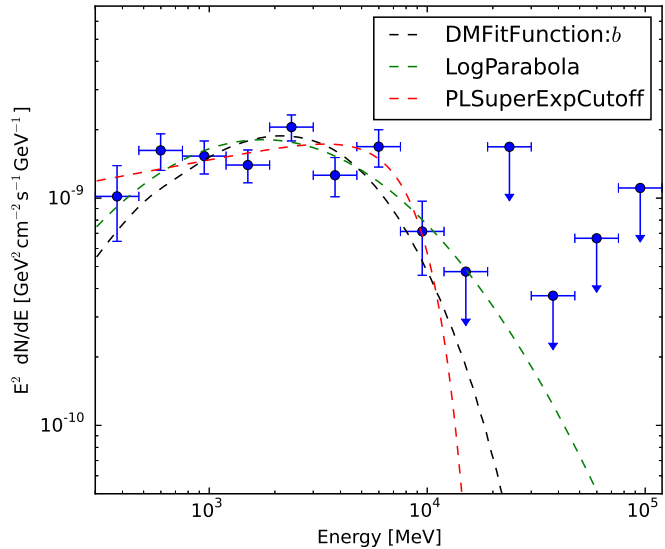}
\caption{The spectral energy distribution (SED) of 3FGL J1924.8-1034 (blue).  Three types of spectra are shown in this figure, including the {\it PLSuperExpCutoff} spectrum in red dashed line, the {\it LogParabola} spectrum in green dashed line and the {\it DMFitFunction} spectrum of dark matter annihilation into $b\bar{b}$ (for a rest mass of 43.5 GeV) in black dashed line.}
\label{sed}
\end{figure}

We change the spectral model to {\it DMFitFunction}\footnote{http://fermi.gsfc.nasa.gov/ssc/data/analysis/scitools/source\_models.html} and perform the likelihood fit with this DM model allowing the DM mass($m_\chi$)and ${J_{\rm factor}}$ to float freely. In the case of dark matter annihilating into $b\bar{b}$ with the annihilation cross section${\rm \langle \sigma v \rangle = 3\times10^{-26}~cm^3/s}$ \cite{steigman}, the data agrees with the DM mass($m_\chi$) of 43.5 GeV (see Fig.\ref{sed}).

To yield the model-independent spectrum of 3FGL J1924.8-1034, we divide the data into 15 evenly spaced logarithmic energy bins from 300 MeV to 300 GeV. We fix the spectral indexes of all sources, and leave their normalizations free in the optimized model.  The likelihood fit is performed in each energy bin and the resulting spectral energy distribution (SED) plotted in Fig.\ref{sed} is similar to that of the Galactic Center excess \cite{zb15gc,calore15gc,huanggc}.

The default setting for the spectral model of 3FGL J1924.8-1034 is {\it PowerLaw} in the 3FGL \cite{3FGL}. According to the derived SED, we consider some other spectral models (i.e., {\it LogParabola} and {\it PLSuperExpCutoff}) to perform the likelihood fit (see Fig.\ref{sed}). Comparing with the {\it DMFitFunction} model, we find ${\rm \Delta \ln{\mathcal{L}=4.2}}$ for the {\it PLSuperExpCutoff} model (only 1.3 for the {\it LogParabola} model, $-27.4$ for the default {\it PowerLaw} model).  However the two spectral models are not nested, so the Wilks theorem doesn't work and a comparison between them is not straightforward \cite{wilk, Algeri16}.  Below we adopt the Akaike Information Criterion (AIC) test \cite{Lande,Akaike} to check which one is better. The AIC is defined as
\begin{equation}
{\rm AIC}= 2k-2\ln{\mathcal{L}}
\end{equation}
where $k$ is the number of parameters of the corresponding model and the model that owns a smaller AIC is the better one.

The {\it PLSuperExpCutoff} model has two more spectral parameters than the {\it DMFitFunction} model, the {\it PowerLaw} model and one more spectral parameter than the LogParabola model. So our results ${\rm AIC_{\rm {\it PLSuperExpCutoff}}}<{\rm AIC_{\rm {\it DMFitFunction}}}<{\rm AIC_{\rm {\it LogParabola}}}< {\rm AIC_{\rm {\it PowerLaw}}}$ indicate that the {\it PLSuperExpCutoff} model provides a better spectral fit than other models. Due to the rather small  ${\rm \Delta \ln{\mathcal{L}}}$, the superiority for the PLSuperExpCutoff model is too small to draw a conclusion.

{\subsection{Nearby Source Confusion}}

Two or more point-like gamma-ray sources in slightly different directions (i.e., the separation between these sources is comparable with or even smaller than the angular resolution of the detector) will be identified as a single spatially-extended source. To check such a possibility we define ${\rm \Delta TS_{\rm 2pts}}$ as twice the increase in the log of the likelihood of the two close point-like sources model compared to that of the one point-like source model.

\begin{equation}
{\rm \Delta TS_{\rm 2pts}}= -2(\ln{\mathcal{L}_{\rm point}}-\ln{\mathcal{L}_{\rm 2pst }}).
\end{equation}

 Placing a point source A at the catalog position of 3FGL J1924.8-1034 ($291.21^{\rm \circ}, -10.58^{\rm \circ}$), we find the best-fit position ($291.29^{\rm \circ}, -10.32^{\rm \circ}$) of the other source B with the help of {\it gtfindsrc}.   Then we fix the source B position at ($291.29^{\rm \circ}, -10.32^{\rm \circ}$) and find the best-fit position of the source A ($291.21^{\rm \circ}, -10.61^{\rm \circ}$). The difference between the best-fit position and the original location of source A is only $0.02 ^{\rm \circ}$, which is well within the LAT PSF.  We notice that the angular separation of these overlapping sources is $0.3^{\rm \circ}$. The number of observed sources in 3FGL at distances $0.3^{\rm \circ}$ is less than the number of sources expected if sources could be detected at arbitrarily small angular separations in Fig.13 of \cite{3FGL}, implying that there could indeed have a nearby source.

To limit the number of new additional degrees-of-freedom, we just take the {\it PowerLaw} as the spectral shape of these two point-like sources. Fitting the spectra of the two point-like sources using the best-fit positions, we get the ${\rm \Delta TS_{\rm 2pts}=58.3}$, which is larger than ${\rm \Delta TS_{\rm ext} =43.8}$ for the stretched extended template, i.e., we have ${\rm \Delta \ln{\mathcal{L}}}$=14.5 for an additional degrees-of-freedom.
However ${\rm \Delta TS_{\rm 2pts}}$ cannot be quantitatively compared with ${\rm \Delta TS_{\rm ext}}$ by using the simple likelihood-ratio test because the models are not nested \cite{wilk, Algeri16}.
Then we also use the AIC test to evaluate which model is significantly better. Compared to the stretched extended template, the two point-like sources model has one more parameters (two more spatial parameters and two more spectral parameters compared to three more extension parameter), we finally have ${\rm AIC_{\rm ext}}>{\rm AIC_{\rm 2pts}}$ (note that ${\rm \Delta TS_{\rm ext}} + 2 < {\rm \Delta TS_{\rm 2pts}}$ and the difference is 12.5), indicating that the data shows slight preference for the two point source hypothesis. However, the difference between these two models is small and the dark matter subhalo hypothesis can not be ruled out.

If 3FGL J1924.8-1034 is actually composed of two point-like sources, one or both of them could have possible counterparts at other wavelengths. So we search them in these four multi-wavelength catalogs, including (1) The Roma BZCAT - 5th edition Multi-frequency Catalogue of Blazars \footnote{http://www.asdc.asi.it/bzcat/}\cite{BZCAT}, (2) CRATES - CRATES Flat-Spectrum Radio Source Catalog \footnote{https://heasarc.gsfc.nasa.gov/W3Browse/radio-catalog/crates.html}\cite{CRATES}, (3) CGRABS - Candidate Gamma-Ray Blazar Survey Source Catalog\footnote{https://heasarc.gsfc.nasa.gov/W3Browse/radio-catalog/cgrabs.html}\cite{CGRABS}, (4)The ATNF Pulsar Catalogue\footnote{http://www.atnf.csiro.au/research/pulsar/psrcat/}\cite{ATNF}. There are two sources found within $1^{\rm \circ}$  from 3FGL J1924.8-1034($291.21^{\rm \circ}, -10.58^{\rm \circ}$), including 5BZBJ1925-1018/CRATES J192503-101834 ($0.28^{\rm \circ}$ away from the 3FGL J1924.8-1034) and CRATES J192627-100555 ($0.63^{\rm \circ}$ away).
Further multi-wavelength studies are needed to establish or rule out the association of these two sources with the possible source B found in our analysis.

\subsection{The Implications}
In the following analysis, we assume that 3FGL J1924.8-1034 is a dark matter subhalo to study its properties. From the fit with the DM model( $b\bar{b}$, ${\rm \langle \sigma v \rangle = 3\times10^{-26}~cm^3/s}$), we obtain that the gamma-ray flux of the subhalo from 300 MeV to 300GeV is ${\rm (3.59\pm0.56)\times10^{-9}ph~cm^{-2}~s^{-1}}$, $m_\chi$ = ${\rm (43.5\pm5.7)}$ GeV and $J_{\rm factor} =(3.45\pm0.54)\times10^{20}~{\rm GeV^2~cm^{-5}}$. In \cite{brun}, ${L= \int_{V_{\rm sub}}\rho^2(r)d^3r=J_{\rm factor}D^{2}}$ is adopted to calculate the subhalo luminosity ($D$ is the distance of the subhalo from the Earth). We can derive that the luminosity-distance ($L-D$) relationship ${L=(7.7\pm1.2)\times10^4 ~ M^2_\odot~{\rm pc}^{-3} {(D/{\rm 1~kpc})}^2}$, which is close to the median distance calculated from Via Lactea-II results in Fig.2 of Brun et al.\cite{brun}. So the dark matter subhalo hypothesis seems viable.

\begin{figure}%[!h]
\includegraphics[width=0.6\columnwidth]{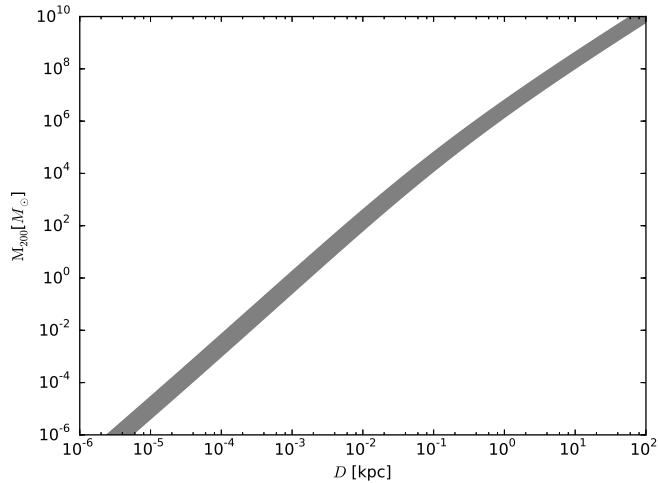}
\caption{ The relationship between the subhalo mass (${\rm M_{200}}$) and the distance ($D$) of the subhalo.  The width of the grey band includes the  effect of scatter in the subhalo's concentration and mass loss for tidal stripping. }
\label{mrb}
\end{figure}

The gamma-ray flux of 3FGL J1924.8-1034 can also be used to constrain the mass and distance of the corresponding dark matter subhalo. The concentration model presented in the \cite{Moline16} which is specifically derived for subhalos is adopted in this calculation. In the Fig.\ref{mrb}, we plot the relationship between the subhalo mass ${\rm M_{200}}$ (before tidal stripping) and the distance $D$ of the corresponding subhalo. The grey band reflects the results for scatter in the subhalo's concentration within a factor of 1.7 and mass loss 90\% due to tidal stripping. However just from the flux, it is not possible to draw further conclusion on the subhalo's mass.

\section{Summary and Discussion}
In this work we have analyzed the spatial distribution, variability and spectrum of the gamma-ray emission from the un-associated source 3FGL J1924.8-1034 . Firstly, for the spatial extension analysis, we take into account three kinds of dark matter distribution templates. In all cases 3FGL J1924.8-1034 prefers a spatially-extended profile over that of a single point-like source (see Fig.\ref{ext}), and the PLE template with an extension radius of ${0.15^{\rm \circ}}$ best describes the data at a high confidence level of $5.4\sigma$. Then we compute the light curve of this source and find no significant evidence for deviation from a constant flux. The spectrum of this source is well described by dark matter annihilation into $b\bar{b}$ with a mass of 43.5 GeV, similar to the value required to interpret the Galactic Center gamma-ray excess. Therefore we conclude that 3FGL J1924.8-1034 is a stable, spatially-extended and unidentified gamma-ray source with a DM-like spectrum.

Dark matter annihilations taking place in some relatively massive and nearby subhalos could appear as stable and spatially-extended gamma-ray sources without detectable counterparts in other bands. Hence, 3FGL J1924.8-1034 maybe an dark matter subhalo candidate though the astrophysical origin can not be convincingly ruled out. In particular, the spectrum of 3FGL J1924.8-1034 is also well described by the spectral shape of {\it PLSuperExpCutoff} which is the spectral model of pulsars. Multiple sources could be misidentified as an extended source. So it is possible that 3FGL J1924.8-1034 is actually made up of two or more gamma-ray sources located closely with each other on the sky. Although the LAT data shows preference for the two nearby point-like gamma-ray sources model, the improvement compared to the stretched extended template is small and the dark matter subhalo hypothesis can not be ruled out.
Supposing that 3FGL J1924.8-1034 is indeed from a dark matter subhalo, we have studied the luminosity-distance ($L-d$) relationship (which agrees with the simulation data, see Section II.D) and the mass-distance ($M_{200}-d$) relationship. Further analysis is thus encouraged to confirm or rule out the dark matter subhalo nature of 3FGL J1924.8-1034.

\begin{acknowledgments}
We thank the referee for the very detailed helpful suggestions and N. Mirabal and J. Graham for their comments.
The data and some analysis tools used in this paper are obtained from the Fermi Science Support Center (FSSC)\footnote{http://fermi.gsfc.nasa.gov/ssc/} provided by NASA Goddard Space Flight Center. This work was supported in part by the National Basic Research Program of China (No. 2013CB837000) and the National Key Program for Research and Development (2016YFA0400200), the National Natural Science Foundation of China under grants No. 11525313 (i.e., the Funds for Distinguished Young Scholars) and No. 11103084, and the 100 Talents program of Chinese Academy of Sciences.

\end{acknowledgments}

%\newpage

\end{document}